\documentclass[reprint,amsmath,amssymb,aps,prd,floatfix]{revtex4-2}

\usepackage{graphicx}
\usepackage{dcolumn}
\usepackage{bm}
\usepackage{amsmath}
\usepackage{enumitem}
\begin{document}

\preprint{APS/123-QED}
\title{Mapping the AMS-02 and CALET Data onto the Source Spectra and Residence Times of Galactic Cosmic Rays}

\author{Ramanath Cowsik}
\email{cowsik@wustl.edu}
\author{Dawson Huth}%
 
\affiliation{%
Washington University in St. Louis,\\
Physics Department and McDonnell Center for the Space Sciences}%

\date{Received 22 September 2023}

\begin{abstract}
The recent measurements of the spectral intensities of cosmic-ray nuclei have suggested that the ratio of Boron to Carbon nuclei, $R(E)$, comprises two components, $R_1(E)$ which carries all of the energy dependence and the other $R_A$, a constant independent of energy per nucleon. This finding supports the earlier theoretical expectations and results of gamma-ray astronomy that one of these components is attributable to spallation in a cocoon like region surrounding the sources and the other in the general interstellar medium before cosmic rays leak away from the Galaxy. A new model-independent way of analyzing cosmic-ray spectra is presented here to shed light on the recent findings: Imposing the single constraint that the source function of B nuclei is minimal we use a cascade of rate equations to map point-by-point the observed cosmic ray spectra of p, B, C, O and Fe on to their source spectra and the two life-times of cosmic rays $\tau_G$ in the Galaxy and $\tau_S$  in the cocoons surrounding the sources. The model independent results show that the appropriate choice is $\tau_S(E)$ is responsible for $R_1(E)$ and the source spectra are power laws with indices of $\sim-2.7$.
\end{abstract}

\maketitle
\section{Introduction}
The low abundances of Li, Be, and B in the Solar System with respect to neighboring elements like C and O, and their significant enhancement in Galactic cosmic rays has provided much insight into the propagation of cosmic rays. Early efforts to interpret this enhancement in a manner consistent with their spectra noted that a steady state was established by a balance between the injection of cosmic rays into the interstellar medium (ISM), where they suffered some spallation and loss of energy, and their subsequent escape from the Galaxy at a rate specified by $\tau$ \cite{Cowsik_1966, Cowsik_1967}. As the cosmic-ray observations became more precise and extended to higher energies, the possibility of spallation and other processes in cocoons of matter surrounding the sources was discussed, especially in the context of the energy-dependence of ratios like B/C \cite{Cowsik_1973, Cowsik_1975}. Subsequently, models explicitly using the diffusion equation to describe the transport of Galactic cosmic rays were progressively developed \cite{Jones_1970, Prishchep_1975, Cowsik_1979, Cesarsky_1980, Moskalenko_1998, Maurin_2001, Strong_2007, Blasi_2009, Cowsik_2010, Maccione_2011, Serpico_2012, Blasi_2013, Burch_2013, Gaggero_2013, Mertsch_2014, Johannesson_2016, Orlando_2018}.


This paper is stimulated by the recent measurements of Adriani et al. \cite{Adriani_O_2020, Adriani_B_2022} of the spectra of Boron, Carbon, other elements and the B/C ratio extending over the energy range 8.4 GeV/$n$ up to 3.8 TeV/$n$ using the CALET instrument aboard the International Space Station. Adriani et al. also plot the B/C ratio as a function of $E/n$ along with the earlier measurements \cite{Engelmann_1990, Swordy_1990, Ahn_2008, Ahn_2009, Panov_2009, Obermeier_2012, Adriani_2014, Aguilar_2018, Aguilar_2021}. As Adriani et al. and others \cite{Adriani_B_2022, Gabici_2023} have noted this observed B/C ratio decreases initially and flattens significantly at higher energies indicating the presence of two separate components one dependent and the other independent of energy, likely arising from propagation of cosmic rays in two regions - cocoons surrounding the sources and the ISM. 

The existence of the two components is confirmed by observations with DAMPE \cite{DAMPE_2022} and were anticipated  in earlier theoretical studies from slightly differing perspectives \cite{Cowsik_1973, Cowsik_1975, Cowsik_2010, Cowsik_2014, Cowsik_2016,  Cowsik_2016b, Evoli_2019, Evoli_2020} and the initial decrease of $R$, the B/C ratio, with energy has been studied using energy-dependent diffusion coefficients \cite{Tomassetti_2012, Aloisio_2015, Johannesson_2016}. The new data provide an important testing ground for the several theoretical suggestions such as energy dependence of diffusion coefficients $D(E)$ \cite{Johannesson_2016, Aloisio_2015, Tomassetti_2012}, which is inversely proportional to the lifetime $\tau(E)$, and to studies investigating the matter traversed in the cocoons from different perspectives \cite{Cowsik_2010, Cowsik_2014, Cowsik_2016, Cowsik_2016b} and \cite{Evoli_2019, Evoli_2020}. We display CALET and DAMPE measurements of the B/C ratio in Figure 1., with an extension to lower energies to accommodate the AMS-02 data \cite{Aguilar_2016}. Even though a flattening of the B/C ratio at high energies was anticipated theoretically it is only the recent data that show the tendency to level-off at high energies. The two components noted by Adriani et al. \cite{Adriani_B_2022} are fit here with two functions:
\begin{equation}
    R_1(E) = R_o E^{-\mu - \delta \ln(E)}
\end{equation}
\begin{equation}
    R_2(E) = R_A = constant,
\end{equation}
with $R_o = 0.267$, $\mu = 0.075$, $\delta = 0.072$, and $R_A = 0.05$, the observed asymptotic value of $R$. \textit{Ab initio}, we do not know which component $R_1(E)$ or $R_A$ is to be attributed to the cocoons, so that the other could be allocated to be generated by spallation effects in the ISM. One of the main objectives of our effort here is to resolve this ambiguity.

In our analysis we map each of the points in the spectra measured by CALET \cite{Adriani_p_2019, Adriani_Fe_2021, Adriani_O_2020, Adriani_B_2022} and AMS-02 \cite{Aguilar_2021} on to the residence times of $\tau_S$ and $\tau_G$, and the injection spectra of p, C, O, and Fe nuclei in the sources. These source spectra then provide a clear choice as to where the energy dependent component of the B/C ratio, $R_1(E)$, is generated. The fits to the observed ratio noted above serve as display-lines in Figure 2.

Our effort in this paper is to map the observed spectra of cosmic-ray nuclei onto their source-spectra and their residence times in the cocoons and subsequently in the ISM. This mapping is carried out with a cascade of rate equations, point-by-point for each energy bin while imposing the single constraint that the source-spectrum of B-nuclei is minimized in conformity with their very low universal abundances.

\begin{figure}
    \centering
    \includegraphics[width=0.45\textwidth]{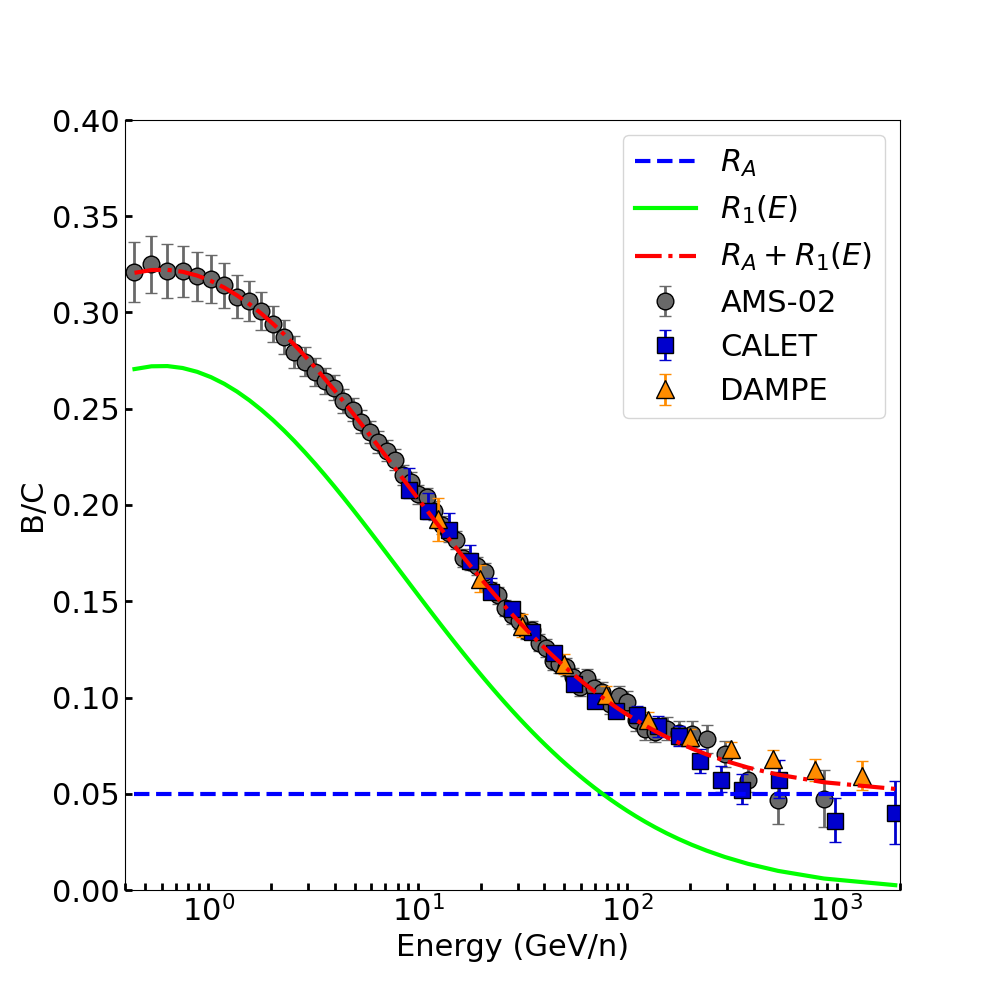}
    \caption{Ratio of the cosmic-ray Boron and Carbon fluxes measured by the CALET \cite{Adriani_B_2022}, AMS-02 \cite{Aguilar_2016}, and DAMPE \cite{DAMPE_2022} instruments. The lines represent energy dependent (solid line) and independent (dashed line) B/C ratios $R_1(E),\ R_A$ and their sum $R_1(E) + R_A$ (chain-dotted line) empirically fit with Equations 1 and 2. The analysis uses each data point individually.}
    \label{B/C_ratio}
\end{figure}

\section{Analysis}
We start the analysis with the standard transport equations; here the spectral intensities of various nuclei are represented as p$(E)$, B$(E)$, C$(E)$, O$(E)$, etc. Our primary focus is in the energy region $>$ 5 GeV/$n$ where the loss of energy due to ionization during traversal of matter and the effects of Solar modulation do not significantly alter the B/C ratio, and the secondary nuclei produced in the spallation process carry away the same $E/n$ as their parent nuclei. The velocity of the nuclei is taken to be $c$. The rate of change of the spectral intensity of a cosmic-ray nucleus $N_j$ during propagation in a region is the sum of the losses due to leakage and spallation of the nucleus, and gain from the production as the spallation product of a heavier nucleus and possible injection by the sources:
\begin{equation}
    \frac{dN_j}{dt} = -N_j\left(\frac{1}{\tau} + c \sigma_{jj}n_H\right) + \sum_{i<j}N_i c \sigma_{ij} n_H + Q_j
\end{equation}
Here the lower value of the subscript, higher is the atomic mass of the nucleus, and $Q_j$ is the source function, $\tau(E)$ is the leakage lifetime of the nucleus in the region under consideration, $n_H$ is the number density of target atoms, $\sigma_{jj}$ and $\sigma_{ij}$ are the total and partial spallation cross sections for the production of $N_j$ from $N_i$, obtained by fitting the data compilations of GALPROP\footnote{{https://galprop.stanford.edu/}} and EXFOR\footnote{{https://www-nds.iaea.org/exfor/}}; see also \cite{Morlino_2020, Maurin_2022}. Under conditions of steady-state, with $\frac{dN_j}{dt} = 0$ this set of equations starting with the heaviest nucleus to the lightest nucleus can be written as a matrix equation
\begin{equation}
    \mathbf{M} \vec{N} = \vec{Q},
\end{equation}
where $\mathbf{M}$ is a lower-triangular matrix, with all the elements above the diagonal being 0 and the diagonal and sub-diagonal elements of the $j$th row are $\mathbf{M}_{jj} = \left(\frac{1}{\tau} + c n_H \sigma_{jj}\right)$ and $\mathbf{M}_{ji} = -c n_H \sigma_{ij}$, $i$ running from 1 to $j-1$. Observed intensities of the cosmic rays are $\vec{N}(E_i)$ and their corresponding source functions are $\vec{Q}(E_i)$. The customary method for solving Equation 4 is to assume $\tau(E)$ and a power-law for $\vec{Q}(E)$ and obtain $\vec{N} = \mathbf{M}^{-1}\vec{Q}$ \footnote{With diffusion equations similar assumptions are needed and one has to sequentially work through the cascade of equations from the heaviest to the lightest nucleus.}. Instead of making all these assumptions, we note that it is adequate to impose a single constraint that the source function $Q_B(E_i)$ be minimized implying that boron nuclei with their negligible universal abundances are not accelerated at the sources and are entirely the products of the spallation of heavier nuclei in cosmic rays. This procedure is carried out by varying $\tau(E_i)$ at each energy $E_i$ to obtain the source function $\vec{Q}(E_i)$ and $\tau(E_i)$ which yields the minimum value.

For propagation in two regions one contained within the other like the ISM and the cocoons around the sources we place the subscripts $G$ and $S$ respectively and write 
\begin{equation}
    \mathbf{M}_G\vec{N}_G = \vec{Q}_G;\  \mathbf{M}_S\vec{N}_S = \vec{Q}_S.
\end{equation}
The source function $\vec{Q}_G$ for cosmic rays in the Galaxy is just the rate at which they leak out of the cocoons surrounding the sources:
\begin{equation}
    \vec{Q}_G = \vec{N}_S/\tau_S \mathrm{\ or\ } \vec{N}_S = \tau_S \vec{Q}_G.
\end{equation}
Combining Equations 5 and 6 we get the basic mapping equation:
\begin{equation}
    \mathbf{M}_S(\tau_S\mathbf{M}_G\vec{N}_G) = \vec{Q}_S.
\end{equation}

\section{Propagation back to the cosmic-ray accelerator reversing the effects of spallation in the ISM and in the cocoon surrounding the source}
The cascade of equations represented by Equation 7 will allow us to map the observed spectra of nuclei $\vec{N}(E)$ to obtain the lifetimes $\tau_G$, $\tau_S$, and the source functions $\vec{Q}(E)$. Note that on the left hand side of Equation 7 the only unknowns are $\tau_G$ and $\tau_S$. The procedure therefore is to take the value $R_A \sim 0.05$, observed at the highest energies, from which we fix $\tau_A$, the constant value of the lifetime needed to generate $R_A$. As we are not propagating all nuclei in this effort we ascribe 20\% \cite{Genolini_2015} of the B flux to cosmic ray nuclei other than the major contributors C and O, namely N, Ne, Mg, S, etc. For a nominal choice of $n_A = 1$ cm$^{-3}$ this procedure yields $\tau_A \sim 1$ Myr and $\lambda_A = c \tau_A n_A m_H \sim 1.6$ g cm$^{-2}$. Keeping $\tau_A$ fixed, $\tau_1$ is varied to minimize $Q_B$ in the sources.



We now encounter two options here: either $\tau_G = \tau_A$ or $\tau_S = \tau_A$, we present the results for these two options one after the other:

\textit{\textbf{Option 1.} $\tau_G = \tau_A = constant;\  \tau_S(E) = \tau_1(E)$ energy dependent.---} Note that in Equation 7 $\tau_S(E)$ occurs both explicitly and implicitly in the terms contained in the matrix $\mathbf{M}_S$. For any given choice of $\tau_S(E_i)$ the equation can be solved readily. The procedure is therefore to fix, as noted before, $\tau_G=\tau_A=1$ Myr $=3.154\times10^{13}$ s with a nominal choice $n_{H,G}\approx1$ cm$^{-3}$ and vary $\tau_S(E_i)$ with a nominal choice of $n_{H,S}\approx200$ cm$^{-3}$, a much larger number, (keeping in mind the possibilities like pre-supernova winds; also, as noted earlier only the product $\tau_S(E_i)n_{H,S}$ matters). This procedure not only determines $\tau_S(E_i)$ and $\vec{Q}_S(E_i)$ but also $\vec{N}_S(E_i)$, and is carried out for all the energy bins where observations are available. The results are displayed in Figures 2 and 3 and are summarized below:

\begin{enumerate}
    \item The $\tau_S(E)$ derived from the analysis, shown in Figure 2., declines steeply with the energy per nucleon as $\sim E^{-1}$ or even steeper at very high energies. As the diffusion coefficient $D$ is inversely proportional to $\tau(E)$, it implies that $D$ is proportional to $p\beta/Z$ (Bohm diffusion) and steepening to $\sim (p\beta/Z)^2$ behavior, in a manner not unlike the diffusion of cosmic rays in the solar wind leading to the modulation of Galactic cosmic rays \cite{Fujii_2003, Komori_2005, Potgieter_2013, Caballero_2019}. Note that \textit{$\mathrm{\tau_S(E)}$ does not have the same energy dependence as of $\mathrm{R_1(E)}$, but is steeper.}
    \item The spectrum of Fe, O, C, and p emerging from the cosmic-ray acceleration, shown in Figure 3., are all nearly identical to one another and over the energy range beyond $\sim 5$ GeV/$n$ fit a power-law with an index of $\sim -2.7$ at higher energies. This indicates that the observed spectra $\vec{N}_G$ are flatter than the source spectra $\vec{Q}_S$ especially at low energies. Also, the C/O ratio in the observed cosmic rays gently decreases at low energies and becomes a constant at high energies. Both these features are caused by greater spallation at lower energies. The falling B/C ratio implies greater spallation at lower energies and less at higher energies. The cross sections are kept independent of energy in the present calculations and we find that the source spectra of all the nuclei fit power laws of the form $\vec{Q}(E) \sim E^{-2.7}$ over the entire energy band from 7 GeV/$n$ to $\sim 2000$ GeV/$n$, and the fit would extend to lower energies when solar modulation effects are corrected for. Splitting the energy domain into two segments, we note that the region below $\sim 50$ GeV/$n$ is slightly steeper and changes to $\sim E^{-2.7}$ at higher energies. These spectral characteristics will change if the cross sections turn out to be dependent on energy. But the changes are expected to be small or negligible.
\end{enumerate}



\begin{figure}
    \centering
    \includegraphics[width=0.45\textwidth]{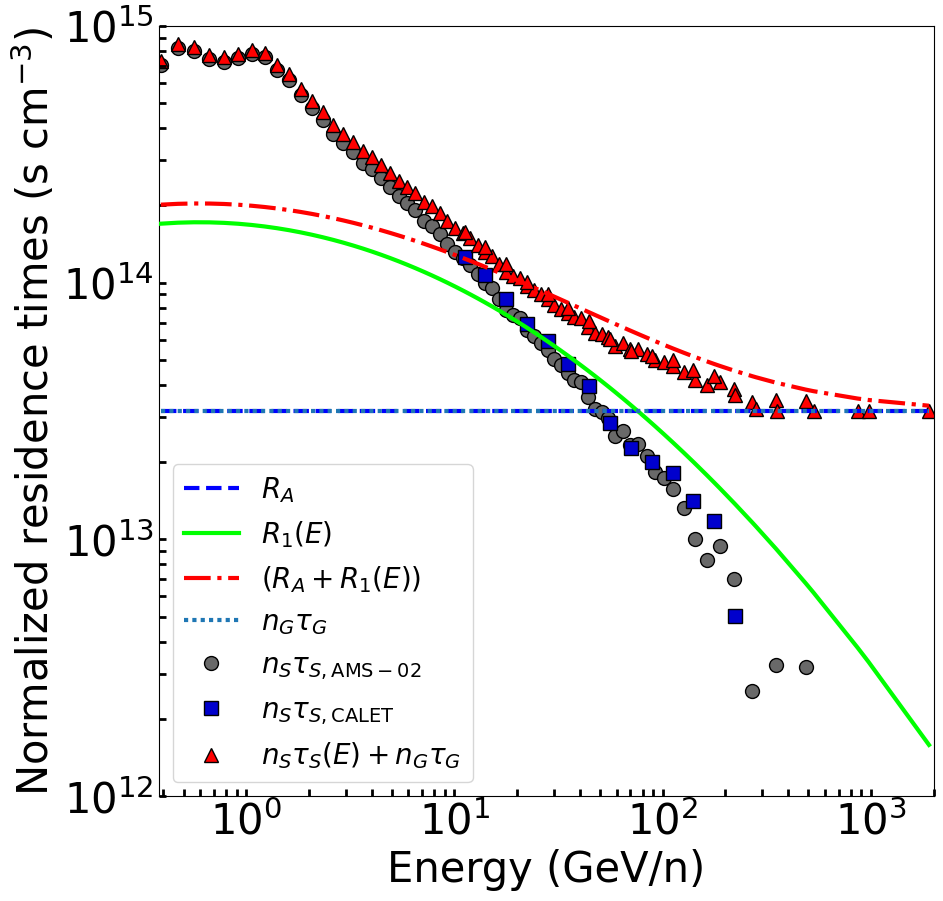}
    \caption{The scaled lifetimes $n_S\tau_S(E)$, $n_G\tau_G$, and $n_S\tau_S(E) + n_G\tau_G$ obtained by mapping the data for Option 1 and the empirical fits to the data on B/C ratios, $R_1(E)$ and $R_A$ (see Equations 2 and 3), which have been scaled to compare their energy dependence. It is clear that one needs $\tau(E)$ to be steeper than $R_1(E)$ to generate the observed B/C ratio. Spallation of B nuclei and their parent nuclei like C and O is the basic cause of this. (Lifetimes are normalized for $n_H = 1$ cm$^{-3}$)}
    \label{residence_times}
\end{figure}

\begin{figure}
    \centering
    \includegraphics[width=0.45\textwidth]{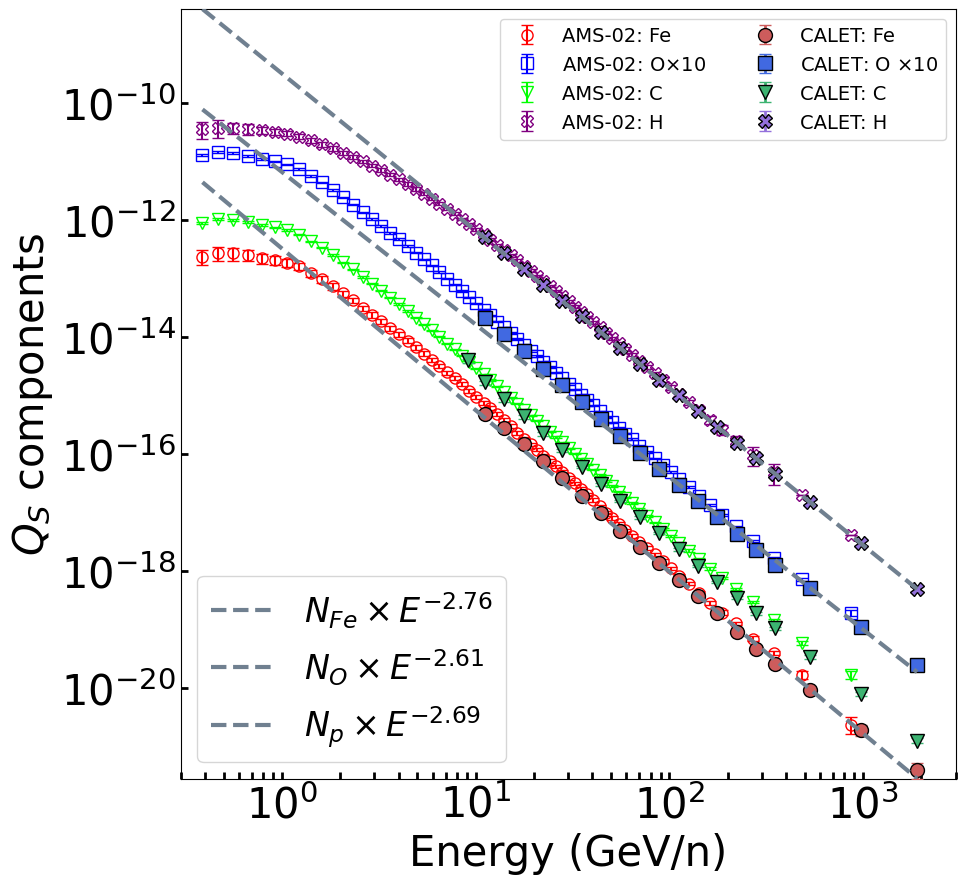}
    \caption{Source spectra obtained by mapping of the CALET and AMS-02 cosmic ray flux data for Option 1. ($\tau_G=\tau_A, \tau_S(E)$). The grey dashed lines represent power laws, $\sim E^{-2.7}$, fit to points at energies greater than 100 GeV/$n$ and normalized to the p, O and Fe source functions obtained by mapping CALET measurements. The spectral index of $Q_C$ is similar to that of $Q_O$. Note that oxygen is multiplied by a factor of 10 for readability. The progressive flattening below $\sim 5$ GeV/$n$ is in part due to modulation effects in the solar wind which were not accounted for.}
    \label{source_components_cocoon}
\end{figure}

\textit{\textbf{Option 2.} $\tau_G(E) = \tau_1(E)\ energy\ dependent, \tau_S = \tau_A = constant$.---}We now reverse the roles of the two regions in our analysis with the source region having a constant residence time $\tau_S$ and the ISM governed by some energy dependent residence time $\tau_G(E)$ as one of the possibilities \cite{Korsmeier_2021, Evoli_2019} suggested by Adriani et al. \cite{Adriani_B_2022}. The prescription is much the same as before except now it is $\tau_G(E)$ being varied to minimize the source function of B. We maintain our previous choice of $n_{H,G} = 1$ and $n_{H,S} = 200$. To ensure cosmic rays traverse the same grammage in the source region as they had in the ISM from Option 1; we take $\tau_S = 3.16\times10^{13}/200 = 1.58\times10^{11}$ s. The source functions, $\vec{Q}_S$, are shown in Figure 4.
\begin{figure}
    \centering
    \includegraphics[width=0.45\textwidth]{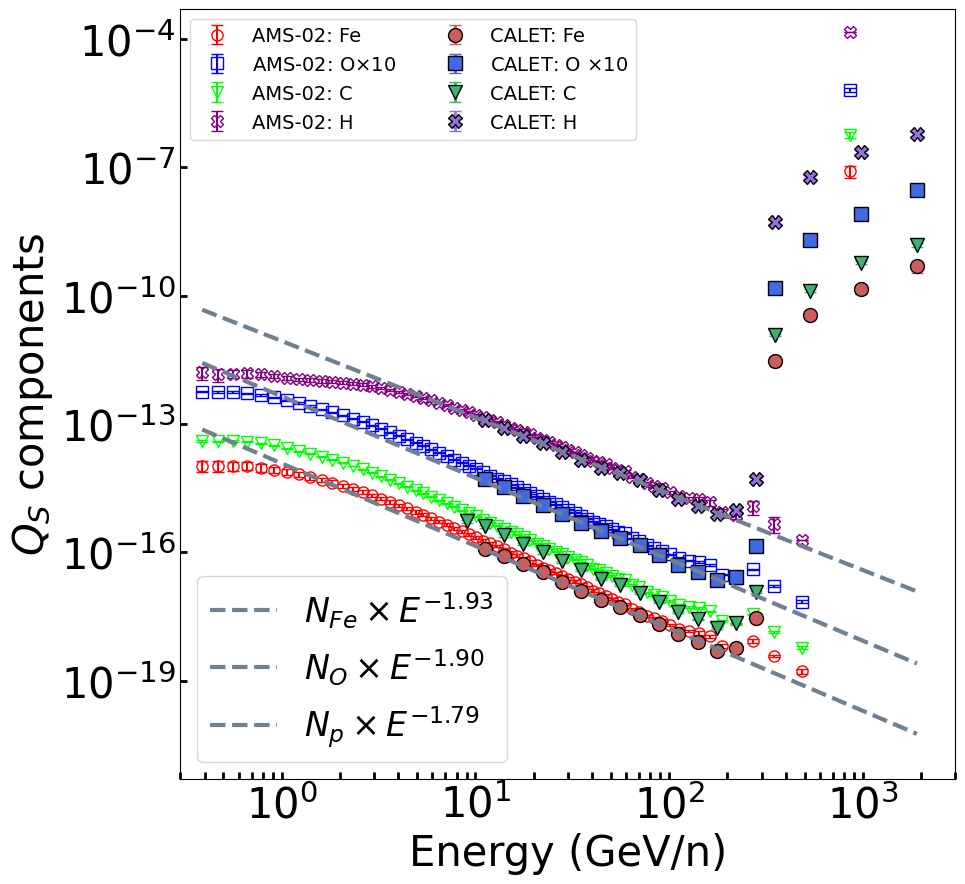}
    \caption{Source spectra obtained for Option 2. ($\tau_G(E), \tau_S=\tau_A$) with energy dependent leakage from the ISM. Unusually flat source spectra $\sim E^{-1.9}$ are needed in this choice. Note that around 200 GeV/$n$ Option 2. becomes problematic. This is because $\tau_G(E)$ has to become rapidly smaller as the B/C ratio approaches the constant $R_A$ value. (This problem at high energies can not be easily remedied but shifted to $\gtrsim 300$ GeV by lowering the value of $\tau_A$ by $\sim 10$\% as allowed by the observational uncertainties, see discussion below.)}
    \label{source_components_galaxy}
\end{figure}
\begin{enumerate}
    \item In contrast with Option 1., here in Option 2. the source spectra are extremely flat and have indices of $\sim -1.9$.
    \item The source spectra become indeterminate at energies above $\sim 200$ GeV/$n$ because $\tau_G(E)$ in the ISM drops sharply towards zero above 200 GeV/$n$ (similar to $\tau_S(E)$ shown in Figure 2). This very small escape time from the Galaxy essentially implies that at high energies as soon as cosmic rays leak out from their source region they also leave the Galaxy. Accordingly, the sources in Option 2. have to put out a very large amount of power in cosmic rays at high energies in order to generate the observed intensities.
\end{enumerate}

\textit{An intuitive understanding} of the mapping results may be gained by noting: a) The lifetimes $\tau$ obtained from the mapping is applicable to all nuclei including protons which suffer minimal depletion due to nuclear interactions; b) When $\tau_S$ becomes very small, it means particles are rapidly injected into the ISM before they suffer nuclear interactions; c) In contrast when $\tau_G$ becomes small, particles leave the Galaxy and the sources have to increase the power to very high values to maintain the observed cosmic-ray intensities; d) Also, the B/C ratio $R_1(E)$, obtained by subtracting $R_A$ from $R(E)$, is much steeper than $R(E)$ especially when $R_1(E)$ approaches $R_A$; e) Finally, $\tau(E)$ corresponding to $R_1(E)$ is steeper than $R_1(E)$ as noted below. Consider just two elements, say B and C, escaping from a single region, say the ISM; the transport equations yield
\begin{align}
    C(E) &= \frac{Q_C(E)\cdot\tau(E)}{1+c\tau(E)\sigma_{CC}n_H} \\
    \mathrm{and}\ R(E) &= \frac{B(E)}{C(E)} = \frac{c\tau(E)\sigma_{CB}n_H}{1+c\tau(E)\sigma_{BB}n_H} \\
    \mathrm{or}\ \tau(E) &= \frac{R(E)}{c n_H \sigma_{BB}(R_x - R(E))}
\end{align}
with $R_x = \frac{\sigma_{CB}}{\sigma_{BB}}\approx0.14$. Note that as $R(E)$ increases towards $R_x$ at low energies, $\tau(E)$ increases rapidly. \textit{Thus $\tau(E)$ is always a steeper function of energy than $R(E)$, which becomes a constant independent of $\tau(E)$ at $R(E) \approx R_x$.}

\section{Discussion of the results} In this paper a new method of analyzing the observed spectra and composition of cosmic rays has been presented: These spectra have been mapped onto $\tau_G$, $\tau_S$ and the source spectra $\vec{Q}$ for the nuclei point-by-point at each observed energy bin. These results are pertinent to the theories of origins and propagation of cosmic rays and the generation of antiparticles like $e^+$ and $\bar{p}$ in the Galaxy \cite{Cowsik_2010, Cowsik_2014, Cowsik_2016, Cowsik_2016b}.
\begin{enumerate}
    \item The main conclusions that emerge are that (i) we have to ascribe the dominant part of the energy-dependent grammage to traversal in the cocoon and hold the leakage lifetime of cosmic rays from the Galaxy a constant or very weakly dependent on energy, (ii) the source spectra are $\vec{Q}(E) \sim E^{-2.7}$, and (iii) the residence time of cosmic rays in the cocoons decrease steeply as $\tau_S(E) \sim E^{-1}$.
    \item We now turn to a discussion of the unexpected behavior of the source spectra displayed in Figure 4. for Option 2. where the energy dependent component of the B/C ratio, $R_1(E)$, is attributed to spallation effects in the ISM, and the energy independent component of B/C arises in the cocoons.
        \begin{enumerate}[label=\alph*)]
            \item Here, the residence time in the Galaxy $\tau_G(E)$ has to steeply decline with energy at energies beyond $\sim 200$ GeV/$n$ where $R_1(E)$ becomes very small at energies where it approaches the constant value $R_A$ from the sources. As a consequence in order to generate the observed spectra the source spectra ought to be very flat or even increasing with energy. As the calculated residence times, $\tau$, are equally applicable to all the different elements, the steepness of $\tau(E)$ becomes especially clear in the case of the spectrum of protons which suffer much smaller reduction through nuclear interactions.
            \item The difficulty addressed above in item a) can not be ameliorated by assuming a small constant lifetime in the ISM $\Delta \tau_A$ at $E > 200$ GeV/$n$. In this situation, $\tau_S$ and $\tau_G(E)$ will both be constant and the injected spectrum will have to rapidly change at $\sim 200$ GeV/$n$ from $\sim E^{-1.9}$ and match the observed spectrum $\sim E^{-2.7}$ with a normalization appropriate to $\Delta \tau_A$.
            \item \textit{This difficulty is avoided in Option 1.} because, $\vec{Q}_G(E) = \vec{N}_S(E)/\tau_S(E)$ and as $\vec{N}_S(E) \sim \vec{Q}_S(E) \tau_S(E)$ at high energies where spallation effects are small (and for protons at almost all energies), $\vec{Q}_G(E)$ sensibly becomes independent of $\tau_S(E)$ at high energies.
        \end{enumerate}
    \item The results presented here are for the energy-independent partial and total cross-sections, $\sigma_{ij}$ and $\sigma_{jj}$. Both the absolute value and the energy dependence of the cross-sections will affect the values and energy dependence of $\tau(E)$ and the spectral indices of the calculated spectra. However, variations are expected to be small. Improved cosmic-ray observations and accurate measurements of the cross-sections up to several 1000 GeV/$n$ are needed especially for fixing the value $R_A$ more accurately.
    \item The available measurements of the B/C ratio only just flatten at the highest energies. For a better determination of $\tau_A$ more observations at higher energies are needed. We note that the ratio $Q_B/Q_C$ values are centered around zero with maximum deviations on the order of 2\% and the main conclusion is robust against changes in $\tau_A$ allowed within the measurement errors.
    \item The constant grammage in the ISM corresponding to $R_A$ is $\sim 1.6$ g cm$^{-2}$, and this is just about adequate to generate the flux of positrons and anti-protons observed in Galactic cosmic rays. The steeply decreasing grammage corresponding to $R_1(E)$, does not contribute significantly to the antiparticle $(\bar{p}\ \mathrm{and}\ e^+)$ fluxes at high energies because of the production-kinematics  \cite{Cowsik_2010, Cowsik_2014, Cowsik_2016, Cowsik_2016b}.
    \item A diverse list of avenues for future investigations are listed below:
        \begin{enumerate}[label=\alph*)]
            \item The steeply declining $\tau_S(E)$ in the cocoons needs to be addressed on the basis of transport theory. In this context Reichherzer et al. \cite{Reichherzer_2022} have shown that the quasi-linear theory of particle transport is valid only when the ratio of the random component in the magnetic fields $\Delta B$ to the background field $B$ is below $0.05$. Observations of Galactic fields yield $\Delta B/B \approx 1$ \cite{Orlando_2013, Adam_2016}; the diffusion coefficient increases rapidly even for much smaller $\Delta B/B$ towards $E^{+1}$ behavior, or equivalently $\tau(E) \sim E^{-1}$. Secondly, Shroer et al. \cite{Schroer_2021, Schroer_2022} have already shown that cocoons around cosmic-ray sources do form due to particle-field interactions. It is also relevant to note that the diffusion coefficient in the interplanetary medium responsible for solar modulation of cosmic rays depends linearly or as a steeper function of rigidity \cite{Fujii_2003, Komori_2005, Potgieter_2013, Caballero_2019}.
            \item Similarly, the near constancy of $\tau_G$ has also to be placed on a theoretical footing - J.Skilling, J.Jokipii, and V.Ptuskin and others \cite{Skilling_1971, Jokipii_1976, Ptuskin_1997} have evoked a convection or advection of cosmic rays in this context. Also Parker \cite{Parker_1966, Parker_1969} has noted that instabilities cause magnetic flux tubes inflated by cosmic-ray pressure to emerge out of the Galactic plane. As this happens the interstellar gas slides down forming clouds and the cosmic rays escape into the intergalactic medium.
            \item Repeating the present analysis by including all nuclides up to and including Fe will provide a consistency check; the spallation cross sections increase as $\sim A^{2/3}$; because of this, spallation effects compete with leakage and become progressively more important in determining the spectral intensities of various heavy nuclei. Furthermore, Sc, Ti, and V get substantial contributions from the spallation of Fe which continues to cascade down to all lighter nuclei. The high abundance of Si, Mg, and Ne will also cascade down contributing to lighter nuclei like F and to a lesser extent Li, Be, and B. Thus the extended study will provide a consistency check on the results obtained with the study of lighter nuclei.
        \end{enumerate}
\end{enumerate}

\section{Concluding remarks}
The observations of the B/C ratio in cosmic rays by the space-borne AMS-02, CALET, and DAMPE instruments show that the ratio decreases initially with energy and tends towards a constant value at energies beyond $\sim 1$ TeV/$n$. This dependence indicates that there exists two sources of production of B nuclei through the spallation of C and heavier nuclei, one that generates the energy dependent part and the other independent of energy. 

These two components are attributable to spallation in two distinct regions, namely, cocoons surrounding the sources of cosmic rays and the other the interstellar medium of the Galaxy. We do not know \textit{ab initio} which region contributes to the energy dependent part of the B/C ratio and which region contributes to the ratio which is independent of energy.

This uncertainty is removed from the analysis presented in this paper. The observed spectra of $p$, B, C, O, and Fe nuclei are mapped at each observed energy on to two lifetimes, one that is energy dependent and the other a constant, and the source spectra of the cosmic ray nuclei. This mapping is effected by adopting a cascade of rate equations and a single assumption that the source strength of B nuclei is minimal in accordance with their universal abundances.

The analysis clearly dictates that the energy dependent part of the spallation occurs in the cocoons surrounding the sources and the constant part arises in the interstellar medium. It also yields the energy dependence of the lifetime and the source spectra of all nuclei to be $\sim E^{-2.7}$.

It is interesting to note that the constant component of the matter traversed in the ISM is $\sim 1.6$ g cm$^{-2}$, which is just about the grammage needed to generate the observed spectral intensities of positrons and antiprotons by the interactions of Galactic cosmic rays \cite{Cowsik_2010, Cowsik_2014, Cowsik_2016, Cowsik_2016b}

\begin{acknowledgments}
    We thank Professor Martin H. Israel for discussions relating to the topics in this paper.
\end{acknowledgments}

\bibliography{bibliography}

\end{document}